\documentstyle[hyperref,twocolumn,epsfig,aps,prl]{revtex}

\begin{document}

\title{The Charge Form Factor of the Neutron from the Reaction 
$^2\vec{\rm H}(\vec{e},e^\prime n)p$}

\author{
\href{http://www.nikhef.nl/user/igorp}{I.~Passchier},$^1$
R.~Alarcon,$^2$
Th.~S.~Bauer,$^{1,3}$
D.~Boersma,$^{1,3}$
J.~F.~J.~van~den~Brand,$^{1,4}$
L.~D.~van~Buuren,$^{1,4}$
H.~J.~Bulten,$^{1,4}$
M.~Ferro-Luzzi,$^{1,4}$
P.~Heimberg,$^{1,4}$\cite{gwu}
D.~W.~Higinbotham,$^{1,5}$
C.~W.~de~Jager,$^{1,6}$
S.~Klous,$^{1,4}$
H.~Kolster,$^{1,4}$
J.~Lang,$^7$
B.~L.~Militsyn,$^1$
D.~Nikolenko,$^8$
G.~J.~L.~Nooren,$^1$
B.~E.~Norum,$^5$
H.~R.~Poolman,$^{1,4}$
I.~Rachek,$^8$
M.~C.~Simani,$^{1,4}$
E.~Six,$^2$
D.~Szczerba,$^7$
H.~de~Vries,$^1$
K.~Wang$^5$ and
Z.-L.~Zhou.$^9$\\
}

\medskip

\address{ 
  $^1$~\href{http://www.nikhef.nl}{National Institute for Nuclear Physics and High Energy
  Physics}, NL-1009 DB Amsterdam, The Netherlands\\
  $^2$~Department of Physics and Astronomy, Arizona State University,
  Tempe, AZ 85287, USA\\
  $^3$~Physics Department, Utrecht University, NL-3508 TA Utrecht,
  The Netherlands\\
  $^4$~\href{http://www.nat.vu.nl}{Department of Physics and Astronomy, Vrije
  Universiteit},
  NL-1081 HV, Amsterdam, The Netherlands\\
  $^5$~Department of Physics, University of Virginia, Charlottesville,
  VA 22901, USA\\
  $^6$~Thomas Jefferson National Accelerator Facility, Newport News, 
  VA 23606, USA\\
  $^7$~Institut f\" ur Teilchenphysik, Eidgen\"ossische Technische Hochschule,
  CH-8093 Z\" urich, Switzerland\\
  $^8$~Budker Institute for Nuclear Physics, Novosibirsk, 630090
  Russian Federation\\
  $^9$~Laboratory for Nuclear Science, Massachusetts Institute of
  Technology, Cambridge, MA 02139, USA}
   
\maketitle

\begin{abstract}
  We report on the first measurement of spin-correlation parameters in
  quasifree electron scattering from vector-polarized deuterium.
  Polarized electrons were injected into an electron storage ring at a
  beam energy of 720~MeV.  A Siberian snake was employed to preserve
  longitudinal polarization at the interaction point.
  Vector-polarized deuterium was produced by an atomic beam source and
  injected into an open-ended cylindrical cell, internal to the
  electron storage ring. The spin correlation parameter $A^V_{ed}$ was
  measured for the reaction $^2 \vec{\rm H}(\vec e,e^\prime n)p$ at a
  four-momentum transfer squared of 0.21~(GeV/$c$)$^2$ from which a
  value for the charge form factor of the neutron was extracted.
\end{abstract}


Although the neutron has no net electric charge, it does have a charge
distribution.  Precise measurements\cite{Kop97} where thermal neutrons
from a nuclear reactor are scattered from atomic electrons indicate
that the neutron has a positive core surrounded by a region of
negative charge.  The actual distribution is described by the charge
form factor $G_E^n$, which enters the cross section for elastic
electron scattering.  It is related to the Fourier transform of the
charge distribution and is generally expressed as a function of $Q^2$,
the square of the four-momentum transfer.  Data on $G_E^n$ are
important for our understanding of the nucleon and are
essential for the interpretation of electromagnetic multipoles of
nuclei, e.g. the deuteron.

Since a practical target of free neutrons is not available,
experimentalists mostly resorted to (quasi)elastic
scattering of electrons from unpolarized deuterium\cite{Lun93,Platchkov}  
to determine this form factor.  The shape of $G_E^n$ as
function of $Q^2$ is relatively well known from high precision elastic
electron-deuteron scattering\cite{Platchkov}.  However, in this case
the cross section is dominated by scattering from the proton and,
moreover, is sensitive to nuclear-structure uncertainties and
reaction-mechanism effects.  Consequently, the absolute scale of
$G_E^n$ still contains a systematic uncertainty of about 50\%.

Many of the aforementioned uncertainties can be significantly reduced
through the measurement of electronuclear spin observables.  The
scattering cross section with both longitudinal polarized electrons
and a polarized target for the $^2 \vec{\rm H}(\vec e,e^\prime
N)$ reaction, can be written as~\cite{Arenh}
\begin{equation}
  S = S_0 \left\{ 1 + P_1^d A^V_d + P_2^d A^T_d + h( A_e + P_1^d
    A^V_{ed} + P_2^d A^T_{ed}) \right\}
\end{equation}
where $S_0$ is the unpolarized cross section, $h$ the polarization of
the electrons, and $P_1^d$ ($P_2^d$) the vector (tensor) polarization
of the target. $A_e$ is the beam analyzing power, $A^{V/T}_d$ the
vector and tensor analyzing powers and $A^{V/T}_{ed}$ the vector and
tensor spin-correlation parameters.  The target analyzing powers and
spin-correlation parameters, depend on the orientation of the target
spin.  The polarization direction of the deuteron is defined by the
angles $\Theta_d$ and $\Phi_d$ in the frame where the $z$-axis is
along the direction of the three-momentum transfer ($\bf{q}$) and the
$y$-axis is defined by the vector product of the incoming and outgoing
electron momenta.  $A^V_{ed}(\Theta_d=90^\circ,\Phi_d=0^\circ)$
contains an interference term, where the effect of the small charge
form factor is amplified by the dominant magnetic form factor
\cite{Scofield,Bill,Arenh}.  At present, there is a worldwide effort
to measure the neutron charge form factor by scattering polarized
electrons from neutrons bound in deuterium and $^3$He nuclei, where
either the target is polarized or the polarization of the ejected
neutron is measured.  Experiments with external beams have been
carried out at Mainz\cite{Meyerhoff} and MIT\cite{Eden,Jones,Thomson}.
In the present paper we describe a measurement performed at NIKHEF
(Amsterdam), which uses a stored polarized electron beam and a
vector-polarized deuterium target.

The experiment was performed with a polarized gas target internal to
the AmPS electron storage ring\cite{AmPS}.  An atomic beam source
(ABS) was used to inject a flux of $3\times 10^{16}$ deuterium atoms/s
(in two hyperfine states) into the feed tube of a cylindrical storage
cell cooled to 75~K. The cell had a diameter of 15~mm and was 60~cm
long.  An electromagnet was used to provide a guide field of 40~mT
over the storage cell which oriented the deuteron polarization axis
perpendicular to $\bf{q}$ in the scattering plane.  A doublet of
steering magnets around the target region compensated for the
deflection of the electron beam by the guide field.  In addition, two
sets of four beam scrapers preceding the internal target were used to
reduce events that originated from beam halo scattering from the cell.
By alternating two high-frequency transitions\cite{ABSRF} in the ABS,
the vector polarization of the target ($P^d_1 = \sqrt{\frac{3}{2}}
(n_+ -n_-)$), with $n_\pm$ the fraction of deuterons with spin
projection $ \pm 1$) was varied every 10 seconds.  Compared to our
previous experiments with tensor-polarized deuterium
\cite{ourT20s,eep,oldtarget}, this target setup resulted in an
increase of the figure of merit by more than one order of magnitude,
with a typical target thickness of $1 \times 10^{14}$
deuterons/cm$^2$.

Polarized electrons were produced by photo-emission from a
strained-layer semiconductor cathode (InGaAsP) prepared to the
negative electron affinity surface state with cesium and
oxygen\cite{PES}. 
The transverse polarization of the
electrons was measured by Mott scattering at 100~keV.
After linear acceleration to 720~MeV the electrons were injected and
stacked in the AmPS storage ring.  In this way, beam currents of more
than 100~mA could be stored, with a life time in excess of 15 minutes.
Every 5 minutes, the remaining electrons were dumped, and the ring was
refilled, after reversal of the electron polarization at the source.
The polarization of the stored electrons was maintained by setting the
spin tune to 0.5 with a strong solenoidal field (using the well-known
{Siberian snake} principle\cite{snake}).  Optimization of the
longitudinal polarization at the interaction point was achieved by
varying the orientation of the spin axis at the source and by
measuring the polarization of the stored electrons with a polarimeter
based on spin-dependent Compton backscattering\cite{CBP}. 

Scattered electrons were detected in the large-acceptance magnetic
spectrometer Bigbite\cite{BB,doug} with a momentum acceptance from 250
to 720~MeV/$c$ and a solid angle of 96~msr (see Fig. 1).  Kinematics
were chosen such that $G_E^n$ was probed near its maximum (as
determined from Ref.  \cite{Platchkov}), resulting in the most
sensitive measurement of $G_E^n$ for a given statistical accuracy.
Consequently, the electron detector was positioned at a central angle
of 40$^\circ$, with an acceptance for the electron scattering angle of
$35^\circ \leq \theta_e \leq 45^\circ$, resulting in a central value
of $Q^2 = 0.21({\rm GeV}/c)^2$.

Neutrons and protons were detected in a time-of-flight (TOF) system
made of two subsequent and identical scintillator arrays.  Each array
consisted of four 160~cm long, 20~cm high, and 20~cm thick plastic
scintillator bars stacked vertically.  Each bar was preceded by two
($\delta E$ and $\Delta E$) plastic scintillators (3 and 10~mm thick,
respectively) of equal length and height, used to identify and/or veto
charged particles.  Each of the 24 scintillators was read out at both
ends to obtain position information along the bars (resolution $\sim 4$~cm) and
good coincidence timing resolution ($\sim 0.5$~ns).  The TOF detector
was positioned at a central angle of 58$^\circ$ and covered a solid
angle of about 250~msr.  Protons with kinetic energies in excess of 40~MeV
were detected with an energy resolution of about 10\%.  The $e^\prime
N$ trigger was formed by a coincidence between the electron arm
trigger and a hit in any one of the eight TOF bars.  By simultaneously
detecting protons and neutrons in the same detector, one can construct
asymmetry ratios for the two reaction channels $^2\vec{\rm H}(\vec
e,e^\prime p)n$ and $^2\vec{\rm H}(\vec e,e^\prime n)p$, in this way
minimizing systematic uncertainties associated with the deuteron
ground-state wave function, absolute beam and target polarizations,
and possible dilution by cell-wall background events.

An experimental asymmetry ($A_{exp}$) can be constructed, via
\begin{equation}
A_{exp}=\frac{N_+ - N_-}{N_++N_-}
\end{equation}
where $N_\pm$ is the number of events that pass the selection
criteria, with $h P_1^d$ either positive or negative. $A_{exp}$ for
the $^2\vec{\rm H}(\vec e,e^\prime p)n$-channel, integrated up to a
missing momentum of 200~MeV/$c$, is shown in Fig.~2 as a function of
time for part of the run.  These data were used to determine the
effective product of beam and target polarization by comparing to the
predictions of the model of Arenh\"ovel \emph{et al.}~\cite{Arenh}.
This advanced, non-relativistic model includes the effects from
final-state interaction, meson-exchange currents, isobar
configurations, and relativistic corrections, and has shown to provide
good descriptions for quasifree proton knockout from tensor-polarized
deuterium\cite{eep}.  Finite acceptance effects were taken into
account with a Monte Carlo code that interpolated the model
predictions between a dense grid of calculations over the full
kinematical range and detector acceptance.  In this way, the effective
product of beam and target polarization (i.e. including the effect of
background events and electron depolarization) was determined to be
$0.42$ with a statistical precision of better than 1\% and a
systematic uncertainty of 3\%, mainly limited by the knowledge of the
proton form factors.

Neutrons were identified by a valid hit in one $E$-scintillator or two
neighboring $E$-scintillators (to allow for events that deposit energy
in two neighboring $E$-scintillators) and no hits in the preceding
($\delta E$ and $\Delta E$) scintillators, which resulted in an 8- to
12-fold veto requirement.  Minimum-ionizing particles and photons were
rejected by a cut on the time of flight, resulting in a clean sample
of neutrons, with only a small proton contamination.  The
spin-correlation parameter was obtained from the experimental
asymmetry by correcting for the contribution of protons misidentified
as neutrons (less than 1\%, as determined from a calibration with the
reaction $^1$H($e,e^\prime p$)), and for the product of beam and
target polarization, as determined from the $^2\vec{\rm H}(\vec
e,e^\prime p)n$ channel.

The main effect of cell wall events is a reduction of the effective
target polarization.  Therefore, the effects largely cancel in the
asymmetry ratio. We have studied the cell wall contribution by
measuring with an empty storage cell.  The background contribution to
the $(e,e^\prime p)n$ and $(e,e^\prime n)p$ channels amounted to 5
$\pm$ 1\%, stable over the entire run. A possible dependence on the
target density was investigated by injecting various fluxes of
unpolarized hydrogen into the cell and measuring quasifree nucleon
knock-out events.  The target density dependence was found to be
negligible at ABS operating conditions.

Figure~3 shows the spin-correlation parameter for the $^2\vec {\rm
  H}(\vec e,e^\prime n)p$ channel as a function of missing momentum.
The data are compared to the predictions of the full model of
Arenh\"ovel \emph{et al.}  \cite{Arenh}, assuming the dipole
parameterization for the magnetic form factor of the neutron and the
Paris nucleon-nucleon ($NN$) potential, folded over the detector
acceptance with our Monte Carlo code for various values of $G_E^n$.
Full model calculations are required for a reliable extraction of
$G_E^n$. This can be seen from Fig.~3, as a Plane Wave Impulse
Approximation (PWIA) calculation for $G_E^n=0$, would result in
$A^V_{ed}(90^\circ,0^\circ)=0$, independent of $p_m$. We extract
$G_E^n(Q^2=0.21 ({\rm GeV}/c)^2) = 0.066 \pm 0.015 \pm 0.004$, where
the first (second) error indicates the statistical (systematic)
uncertainty. The systematic error is mainly due to the uncertainty in
the correction for misidentified protons and the orientation of the
holding field (thus the contribution of the spin-correlation parameter
$A^V_{ed}(0^\circ,0^\circ)$ to our experimental asymmetry).

We have investigated the influence of the $NN$ potential on the
calculated spin-correlation parameters using Arenh\"ovel's full
treatment.  The results for $A^V_{ed}(90^\circ,0^\circ)$ using the
Paris, Bonn, Nijmegen, and Argonne V$_{14}$ $NN$ potential differ by
less than 5\% for missing momenta below 200 MeV/$c$.

In Fig.~4 we compare our experimental result to other data obtained
with spin-dependent electron scattering. Note that all other data have
been obtained from a comparison to PWIA predictions and thus without
taking into account reaction mechanism effects.  The figure also shows
the results from Ref.~\cite{Platchkov}, where the upper and lower
boundaries of the `shaded' area correspond to their result obtained
with the Nijmegen and Reid Soft Core potentials, respectively.  It is
seen that our result favors their extraction of $G_E^n$ which uses the
Nijmegen potential. By comparison to the predictions of the QCD-VM
model by Gari and Kr\" umpelmann, with \cite{GK1} and without
\cite{GK2} the inclusion of the coupling of the $\phi$-meson to the
nucleon (which these authors identify with the effect of strangeness
in the neutron), our datum favors the prediction without strangeness
in the neutron included.

In summary, we presented the first measurement of the sideways
spin-correlation parameter $A^V_{ed}(90^\circ,0^\circ)$ in quasifree
electron-deuteron scattering from which we extract the neutron charge
form factor at $Q^2 = 0.21$ (GeV/$c$)$^2$. When combined with the
known value and slope \cite{Kop97} at $Q^2 = 0$ (GeV/$c$)$^2$ and the
elastic electron-deuteron scattering data from Ref. \cite{Platchkov},
this result puts strong constraints on $G_E^n$ up to $Q^2 = 0.7$
(GeV/$c$)$^2$.

We would like to thank the NIKHEF and Vrije Universiteit
technical groups for their outstanding support and Prof. H. Arenh\"
ovel for providing the calculations.  This work was supported in part
by the Stichting voor Fundamenteel Onderzoek der Materie (FOM), which
is financially supported by the Nederlandse Organisatie voor
Wetenschappelijk Onderzoek (NWO), the National Science Foundation
under Grants No.  PHY-9504847 (Arizona State Univ.), US Department of
Energy under Grant No. DE-FG02-97ER41025 (Univ. of Virginia) and the
Swiss National Foundation.

\begin{figure}
  \epsfig{figure=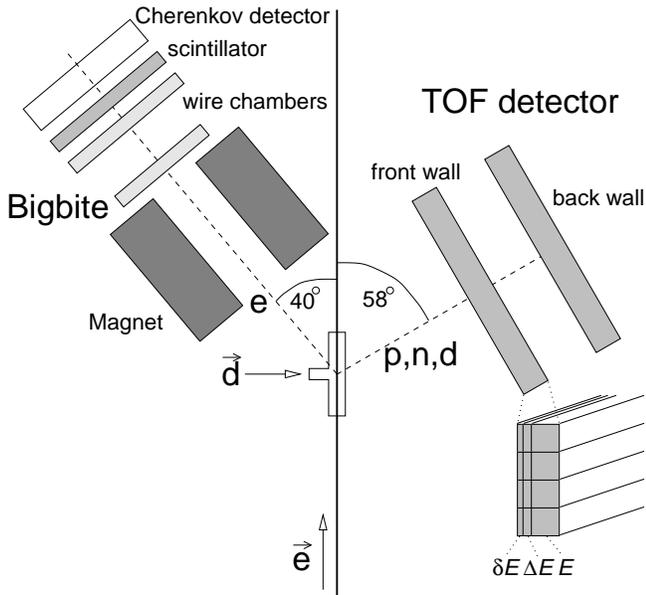,width=3.375in}
\caption{Layout of the detector setup. The electron
  spectrometer consists of a 1~Tm magnet, two drift chambers of four
  planes each, a scintillator and a \v Cerenkov detector.  The time of
  flight system consists of two identical walls of four
  $E$-scintillators preceded by two ($\delta E$ and $\Delta E$) veto
  scintillators.}
\label{fig1}
\end{figure}
\begin{figure}
  \epsfig{figure=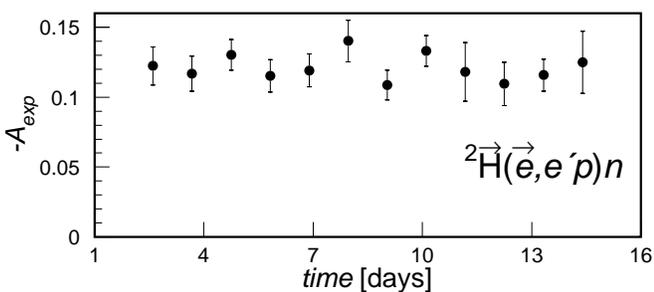,width=3.375in}
\caption{Asymmetry for the reaction $^2\vec{\rm H}(\vec e,e^\prime
  p)n$ integrated up to a missing momentum of 200~MeV/$c$ versus time
  for a two-week period.}
\label{fig2}
\end{figure}
\begin{figure}
  \epsfig{figure=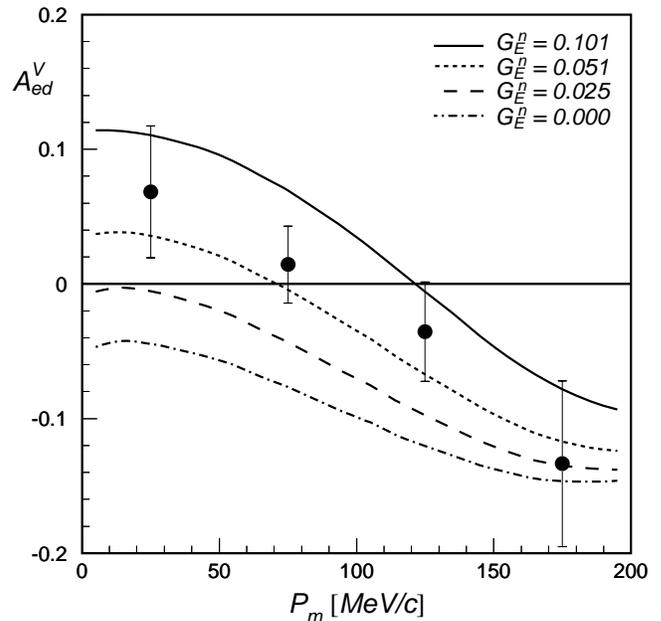,width=3.375in}
  \caption{Data and theoretical predictions for the sideways asymmetry
    $A^V_{ed}(90^\circ,0^\circ)$ versus missing momentum for the $^2
    \vec{\rm H}(\vec e,e^\prime n)p$ reaction. The curves represent
    the results of the full model calculations of Arenh\"ovel \emph{et
      al.} assuming $G_E^n$ equals 0, 0.5, 1.0 and 2.0 times the
    Galster parameterization\protect\cite{Galster}, which results in
    the values of $G_E^n$ at $Q^2 = 0.21 ({\rm GeV}/c)^2$ as shown in
    the legend. A PWIA calculation for $G_E^n = 0$ would result in
    $A^V_{ed}(90^\circ,0^\circ)=0$, independent of $p_m$.}
  \label{fig3}
\end{figure}
\begin{figure}
  \epsfig{figure=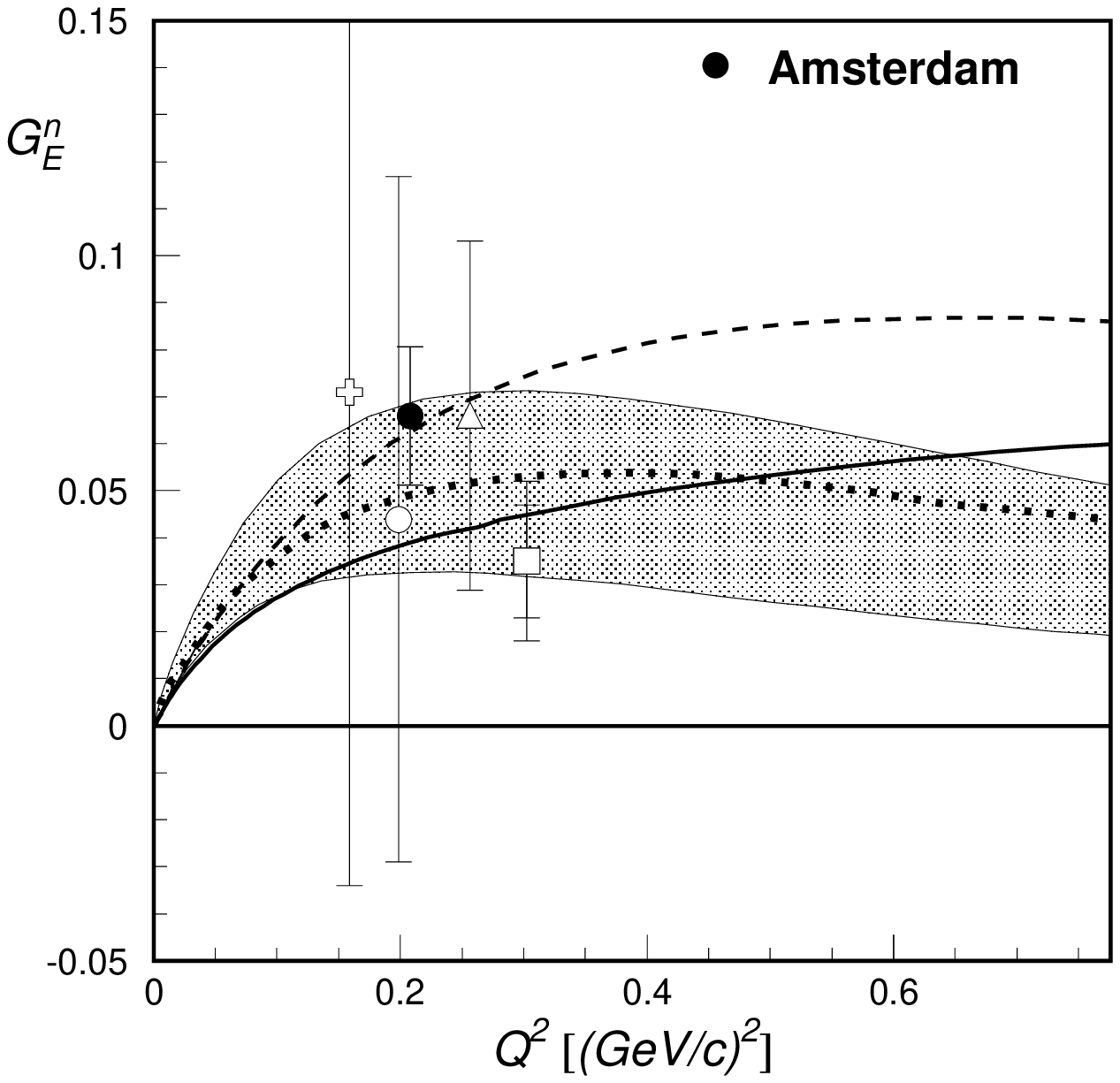,width=3.375in}
  \caption{Data and theoretical predictions for the charge form factor
    of the neutron as a function of four-momentum transfer. The solid
    circle shows our result. The cross and open circle represent the
    results from inclusive measurements performed at
    MIT-Bates\protect\cite{Jones,Thomson}, where an external polarized
    electron beam was scattered from polarized $^3$He. The square
    represents the datum for an electron-neutron coincidence
    measurement with polarized $^3$He obtained at
    Mainz\protect\cite{Meyerhoff}, whereas the triangle represents the
    result of a $^2{\rm H}( \vec e ,e^\prime\vec n)p$
    polarization-transfer experiment at Bates \protect\cite{Eden}. The
    shaded area indicates the systematic uncertainty from the
    unpolarized data by Platchkov \emph{et
      al.}\protect\cite{Platchkov}.  The dotted curve shows the result
    of Galster {\sl et al.}\protect\cite{Galster}, while the solid and
    dashed curves represent the theoretical predictions of Gari and
    Kr\"umpelmann\protect \cite{GK1,GK2} with and without inclusion of
    $\phi$-nucleon coupling, respectively.}  \label{fig4}
\end{figure}

\end{document}